\documentclass[11pt]{article}
\usepackage[centertags]{amsmath}
\usepackage{amsfonts}
\usepackage{amssymb}

\newcommand{\be}{\begin{equation}}
\newcommand{\ee}{\end{equation}}
\newcommand{\bea}{\begin{eqnarray}}
\newcommand{\eea}{\end{eqnarray}}
\begin{document}

\title{%
Doubly Special Relativity at the age of six}
\author{%
J.\ Kowalski--Glikman\thanks{Institute for Theoretical Physics,
University of Wroclaw,   Poland;  {\tt
jurekk@ift.uni.wroc.pl}}%
} \maketitle

\begin{abstract}
The current status of
Doubly Special Relativity research program is shortly presented.

I dedicate this paper to my teacher and friend Professor Jerzy Lukierski on occasion of his seventieth birthday.
\end{abstract}

\section{Introduction}

The definition of  Doubly Special Relativity (DSR) (see
\cite{Amelino-Camelia:2000ge}, \cite{Amelino-Camelia:2000mn} for
the original proposal and \cite{Kowalski-Glikman:2004qa} for review)
is deceptively simple. Recall that Special Relativity is based on
two postulates: Relativity Principle for inertial observers and 
existence of a single observer independent scale associated with
velocity of light. In this DSR replaces the second postulate by
assuming existence of {\em two} observer independent scales: the old
one of velocity plus the scale of mass (or of momentum, or of
energy). That's all.

As any good idea, the DSR proposal has raised a lot of deep and
interesting questions. Among them the immediate ones are: is it
possible to construct an example of such a theory? if so, what makes
us sure that we have to do with something really new and not with,
so to say, Special Relativity in disguise? if so again, how unique
is the construction? is DSR fundamental or somehow emergent as a
particular limit of some more fundamental theory? and last but not
least, are there any deviations from Special Relativity predicted by
DSR, which can be observed in experiments in a foreseeable future? 

After six years of investigations we have now a rather clear picture
of DSR and, at least in the framework based on non-commutative
geometry and quantum algebras of spacetime symmetries we know the
answers to most of these questions. Before turning to describing the
developments, and the current state of art in DSR, let me emphasize that
the progress, which we witnessed in the last years would be
impossible had it not for the contribution of J.\ Lukierski and his
constant stressing the role that quantum symmetries must play in any
attempt to go beyond the standard Special Relativity\footnote{There
are, of course other approaches to the DSR idea. For example Joao
Magueijo is trying to formulate a theory based on DSR principles
without involving mathematics of quantum symmetries. Since this
approach is based on quite different physics and mathematics 
I will not describe it here; I refer the reader to the original
paper \cite{Magueijo:2006qd} and references therein.}.

In what follows I will concentrate on the aspect of DSR that in my
view is most important: quantum $\kappa$-Poincar\`e algebra with
associated non-commutative $\kappa$-Minkowski space.

\section{Deformed $\kappa$ stuff}

The acronym DSR is deciphered by some to mean Deformed Special
Relativity. This name captures the key property of DSR, being a
deformations of Special Relativity, with the scale $\kappa$, of
dimension of mass, identified usually with Planck mass, being the
deformation scale. An example of one such possible deformation,
which plays a role of the main building block of the
``noncommutative'' approach to DSR is the $\kappa$-Poincar\`e algebra
proposed first in the papers by J.\ Lukierski and his
collaborators \cite{Lukierski:1991pn}, \cite{Lukierski:1992dt} and
presented in the final, so-called bicrossproduct form in
\cite{Lukierski:1993wx} and \cite{Majid:1994cy}. In the
bicrossproduct basis the Lorentz subalgebra of $\kappa$-Poincar\'e
algebra, generated by rotations $M_i$ and boosts $N_i$ is not
deformed (purely classical)
$$
[M_i, M_j] = i\, \epsilon_{ijk} M_k, \quad [M_i, N_j] = i\,
\epsilon_{ijk} N_k,
$$
\begin{equation}\label{1}
  [N_i, N_j] = -i\, \epsilon_{ijk} M_k.
\end{equation}
and the deformation is present only in the way the boost act on
commutative momenta
\begin{equation}\label{2}
  [M_i, k_j] = i\, \epsilon_{ijk} k_k, \quad [M_i, \omega] =0
\end{equation}
\begin{equation}\label{3}
   \left[N_{i}, {k}_{j}\right] = i\,  \delta_{ij}
 \left( \frac12 \left(
 1 -e^{-2{\omega}}
\right) + {{\mathbf{k}^2}\over 2}  \right) - i\, k_{i}k_{j} ,
\end{equation}
and
\begin{equation}\label{4}
  \left[N_{i},\omega\right] = i\, k_{i}.
\end{equation}
For completeness let me recall the Casimir of this algebra
\begin{equation}\label{5}
 {\cal C}(\omega, \mathbf{k}) = \left(2\sinh{\omega/2}\right)^2 - {\mathbf{k}^2}\, e^{\omega}
\end{equation}
Naively one would be tempting to conclude that this deformation of SR leads to
energy-depending speed of light, but let me not jump onto this
conclusion too easily and go on. The immediate problem with this
algebra is the question if it is not just SR in disguise, with 
funny variables replacing physical momenta. This would clearly be
the case had it not for additional structures of quantum
$\kappa$-Poincar\`e algebra.

These additional structures are co-product $\Delta$ (roughly telling how generators of the algebra act on products and thus measuring deviation from the standard Leibniz rule) and antipode $S$ (being generalized minus). These structures arise quite naturally \cite{Kowalski-Glikman:2004tz} if one realize that $\kappa$-Poincar\`e algebra can be understood from the point of view of group theory on (part of) de Sitter space. With these structures $\kappa$-Poincar\`e algebra is ``stable'' in a sense that it cannot be transformed to the standard Poincar\`e algebra\footnote{For the standard Poincar\`e algebra both co-product and antipode are trivial in the sense that the former just reflects the Leibniz rule, while the latter if the ordinary minus.} by change of variables: one can easily find new variables in terms of which the algebra (\ref{1})--(\ref{4}) takes the standard, linear form, but in  these new variables the coproduct and antipode are still nontrivial. In general, $\kappa$-Poincar\`e algebra is nontrivial exactly because it cannot be turned into the standard one by any change of variables.

An important consequence of the nontrivial coproduct is that spacetime of DSR is non-commutative \cite{Lukierski:1993wx}, \cite{Majid:1994cy}. The algebra of non-commutative coordinates reads
\begin{equation}\label{6}
   [\hat x_0, \hat x_i] =- \frac{i}{\kappa}\, \hat x_i
\end{equation}
and remarkably the form of this noncommutativity does not depend on the particular basis of $\kappa$-Poincar\`e algebra one starts with.
To derive this result one noticed that spacetime is dual to momentum space and uses the fact that the action of any generators of $\kappa$-Poincar\`e algebra on a product is governed by coproduct.

Given $\kappa$-Minkowski space, an obvious thing to do is to try to construct field theory on it and investigate its properties. To do that one would first need to find out how plane waves on such space look like \cite{Amelino-Camelia:1999pm}. Because of noncommutativity, we must order the plane waves somehow and I choose the ordering ``time to the right'', i.e.
\begin{equation}\label{9}
\hat{e}_k \equiv   e^{i \mathbf{k}\, {\mathbf{\hat x}}}\, e^{-ik_0 \hat x_0}.
\end{equation}
Interestingly enough these plane waves can be understood as arising from the Iwasawa decomposition of the $SO(4,1)$ group. Namely the noncommutative positions of $\kappa$-Minkowski space (\ref{6}) can be regarded as Lie algebra $b_\kappa$ of Lie group $B_\kappa$ that I will call Borel algebra and group, respectively. This group arises naturally in Iwasawa decomposition of the $SO(4,1)$ into ``Lorentz'' and ``translational'' parts, and can be therefore identified with (a
portion of) de Sitter space of momenta. Thus the ordered plane waves above can be equivalently regarded as Borel group elements. Accordingly,
the composition of plane waves, usually understood in terms of the
coproduct of $\kappa$-Poincar\'e algebra, can be regarded as a
simple group elements product. Explicitly
\begin{equation}\label{10}
 \hat{e}_k \hat{e}_p = \hat{e}_{kp}, \quad  kp\equiv (k_0 + p_0, k_i +
e^{-\frac{k_0}{\kappa}}\, p_i)
\end{equation}
Note that Borel group can be
coordinatized by labels $k$ in the plane waves. These coordinates
correspond to the ``cosmological coordinates'' on (the portion of)
de Sitter space of momenta \cite{Kowalski-Glikman:2003we}.

\section{Calculus and field theory}

Knowing what plane waves are, and therefore knowing how to describe quanta of definite energy and momentum, we can try now to combine them into fields. Before doing so I must first explain briefly how to define calculus on $\kappa$-Minkowski space, which is necessary to write down action and field equations.

As for the differential calculus, a natural requirement is to demand that differentiation is Lorentz covariant in some well defined sense, called by mathematicians ``bicovariant differential calculus.'' For $\kappa$-Minkowski space such, essentially unique calculus was derived in \cite{5dcalc1}. For any time to the right ordered function one defines the differential
\begin{equation}\label{11}
    d f(\mathbf{\hat x}, \hat x_0) =  dx^A\, \hat \partial_A \, f( \mathbf{\hat x}, \hat x_0)
\end{equation}
with {\em five}, instead of four, dimensional basis of one forms $dx^A$. Among them four $dx^\mu$ transform linearly under Lorentz generators, as in the standard case, while $dx^4$ is a Lorentz scalar. It should be stressed that in the case of $\kappa$-Minkowski space one cannot do any better than that: although a four dimensional calculus on this space exists it is not Lorentz covariant and therefore does not seem to be very useful in the case of a theory, like DSR, for which Lorentz symmetry is the key requirement (which does not mean that one cannot get very interesting results using this calculus, see \cite{Agostini:2006nc}, for example.)

Note that in (\ref{11}) a derivative $\hat \partial_A$ appear. The action of this derivative on a plane wave $e_k$ labeled with $k$ gives
\begin{equation}\label{12}
  \hat{\partial}_\mu e_k(\hat{x})=P_\mu e_k(\hat{x}),\quad \hat{\partial}_4e_k(\hat{x})=(P_4+\kappa)e_k(\hat{x})
\end{equation}
where  $P_A(k_0, \mathbf{k})$ are given by
\begin{eqnarray}
 {P_0}(k_0, \mathbf{k}) &=& \kappa \sinh
{\frac{k_0}{\kappa}} + \frac{\mathbf{k}^2}{2\kappa}\,
e^{ \frac{k_0}{\kappa}} \nonumber\\
 P_i(k_0, \mathbf{k}) &=&   k_i \, e^{ \frac{k_0}{\kappa}} \nonumber\\
 {P_4}(k_0, \mathbf{k}) &=&
  -\kappa\cosh {\frac{k_0}{\kappa}}  + \frac{\mathbf{k}^2}{2\kappa} \,
e^{ \frac{k_0}{\kappa}}   \label{13}
\end{eqnarray}

The only remaining ingredient of the calculus is integration, which is defined quite naturally as
$$
\int d^4\hat x\, e_k(\hat x) = \delta^4(k)
$$

Having all these ingredients we can formulate free scalar field theory. In the case of massive complex field, for which
\begin{equation}\label{14}
    \hat \phi(\hat{x})=\frac{1}{(2\pi)^{4}}\int d\mu \phi(k)\,e_k(\hat{x})
\end{equation}
\begin{equation}\label{15}
     \hat \phi^\dagger(\hat{x})=\frac{1}{(2\pi)^{4}}\int d\mu \overline{\phi(k)}\,e_{S(k)}(\hat{x})
\end{equation}
with
\begin{equation}\label{16}
   S(\hat{k}_i)=-\hat{k}_ie^{\hat{k}_0/\kappa} ,\,\,\, S(\hat{k}_0)=-\hat{k}_0
\end{equation}
the action has the form
\begin{equation}\label{17}
    S = \frac12\int d^4\hat x\, \hat \partial_\mu \hat \phi^\dagger \hat \partial^\mu \hat \phi + m^2\hat \phi^\dagger\hat \phi
\end{equation}

\section{Recent developments}

Let me now turn to short review of the developments that took place in the recent months.
\newline

{\bf 1. Noether charges}. In the previous section we did not discuss the physical meaning of the objects denoted by $k$. Are they physical momenta carried by elementary quanta, or just labels? This question was recently analyzed in depth in \cite{Agostini:2006nc}. The  conclusion if this paper (quite obvious and natural {\em a posteriori}) is that the physical meaning can be only associated with Noether charges and not with labels of the plane wave directly. Thus physical momentum is a Noether charge associated with physical translational symmetry and not a $k$ label of plane wave $e_k$ (the two coincide for a single quantum in the standard case.) I believe that closer investigation of this question will lead to the conclusion that to large extend the freedom of choosing DSR basis, or plane wave labels, is a kind of coordinate transformation in momentum space, with no physical consequences.\newline

{\bf 2. Star product and Minkowski spacetime DSR field theory.} The field theory (\ref{17}) is pretty elegant and compact, but the fact that it is defined on noncommutative $\kappa$-Minkowski space makes it hard to discuss physical questions. Therefore it would be handy to have an equivalent formulation of the same theory on the standard Minkowski spacetime. This can be actually done and the result is surprisingly simple \cite{Freidel:2006gc}. One defines the star product on Minkowski spacetime to satisfy two requirements: the product of two functions on $\kappa$-Minkowski space equals to the star product of the corresponding ones on Minkowski space-time, and bicovariant derivative $\hat \partial_\mu$ (\ref{11}) on $\kappa$-Minkowski goes to the standard partial derivative $\partial_\mu$ on  Minkowski. With this two ingredients one can calculate the action on Minkowski spacetime equivalent to (\ref{17}) and the result is
\begin{eqnarray}
S&=&\int \mathrm{d}^4x\, \frac{1}{2} (\partial_{\mu} \phi)^\dagger \star (\partial^{\mu} \phi)(x) + \frac{m^2}{2}\, \phi^\dagger\star \phi(x)\label{48}\\
&=&\int \mathrm{d}^4x\, \frac{1}{2} (\partial_{\mu} \phi)^*
(1-\partial_{4}) (\partial^{\mu} \phi)(x) + \frac{m^2}{2}\,
\phi^*(1-\partial_{4}) \phi(x)\label{49}
\end{eqnarray}
where $*$ denotes
the complex conjugation and
\begin{equation} (1-\partial_4) = \sqrt{
1 +\Box}, \quad \Box= ( - \partial_0^2 +
\partial_i\partial^i)
\end{equation}
is the differential operator arising in the bicovariant differential
calculus \cite{5dcalc1}. This action is, of course, manifestly invariant under action of Poincar\`e transformations. Looking at dispersion relation following from this action, one immediately sees that there is no effect of energy dependent speed of light, at least not on the kinematical level (it is not clear at present what is going to happen when interactions are switched on.)\newline

{\bf 3. DSR from gravity.} Another exciting direction of investigations is to find out if DSR can be understood as a particular limit of gravity, coupled to point particles and fields, in the flat space limit, when dynamical degrees of freedom of gravity can be neglected. This question was one of the themes of the recent paper \cite{Kowalski-Glikman:2006mu} (based on seminal paper by Freidel and Starodubtsev \cite{Freidel:2005ak} and our paper \cite{Freidel:2006hv}.) Starting from the results concerning new gravitational perturbation theory investigated there, we speculate that in the relevant DSR limit gravity with particles can be described as effectively three dimensional, boundary theory, described by $SO(4,1)$ Chern-Simon topological theory living on three dimensional spacetime, whose space has punctures labeled by $SO(4,1)$ charges corresponding to particles. Quantization of such theory would lead to emergence of quantum algebra $SO_q(4,1)$, which in the limit of vanishing cosmological constant may contract down to $\kappa$-Poincare algebra \cite{Lukierski:1992dt}, and thus to a theory of DSR type. Of course, all the technical steps are still to be done, but the path from quantum gravity to DSR seems to be, although not completely clear, at least well marked out.

\section{Questions \& answers}

Let me conclude with short answers to the the questions posed at the beginning of this paper\footnote{Beware! These answers reflect my personal understanding and views, and may not be shared by other experts!}

Is it
possible to construct an example of such a theory? Yes, and example being 2+1 gravity coupled to particles and/or fields \cite{Amelino-Camelia:2003xp}, \cite{Freidel:2003sp}, \cite{Freidel:2005me}. The situation in 3+1 dimensions is not clear, but preliminary results are really encouraging.

What makes
us sure that we have to do with something really new and not with,
so to say, Special Relativity in disguise? It is obvious, topological degrees of freedom of gravity do the job.

How unique
is the construction? Assuming standard formulation of gravity in 2+1 as Chern-Simon theory it is unique.

Is DSR fundamental or somehow emergent as a
particular limit of some more fundamental theory? It is clearly a limit of gravity. The question is what is the correct limit in 3+1 dimensions, SR, DSR, or perhaps there are two different limits in different regimes.

Are there any deviations from Special Relativity predicted by
DSR, which can be observed in experiments in a foreseeable future? The field theory presented above strongly suggest that there should be no such observable effects in cosmic rays (the predictions of DSR for both GLAST and Pierre Auger signals seem to be effectively zero.) But there might be interesting deviations for large quantum systems of energies close to the Planck one, see \cite{Magueijo:2006qd} for the concrete proposal.

\section*{Acknowledgements}

This work  is partially supported by the grants KBN 1 P03B 01828.

\end{document}